\documentclass{article}
\usepackage[left=3.00cm, right=3.00cm, top=2.00cm, bottom=2.00cm]{geometry}
\usepackage{authblk}
\usepackage{epsfig}
 \usepackage{caption}
\usepackage{algorithm}
\usepackage{algorithmic}
\usepackage{latexsym}
\usepackage{amsmath}
\usepackage{amsfonts}
\usepackage{amsthm}
\usepackage{graphicx}
\usepackage[normalem]{ulem}
\usepackage{array}
\usepackage{enumerate}
\usepackage{color}\usepackage[dvipsnames]{xcolor}
\usepackage{tikz}\usetikzlibrary{patterns}
\usepackage{exscale}
\usetikzlibrary{decorations.markings}

\newtheorem{lemma}{Lemma}

\newtheorem{corollary}{Corollary}
\newtheorem{theorem}{Theorem}

\newtheorem{prop}{Proposition}

\newcommand{\Av}{\mathtt{Av}}
\newcommand{\Sort}{\mathrm{Sort}}
\newcommand{\out}{\mathrm{out}}

\newcommand{\comm}[1]{#1}
\def\a{0cm} 
\def\b{1cm} \def\B{1.5cm}
\def\c{2cm} \def\C{2.5cm}
\def\d{3cm} 
\def\e{4cm} \def\E{4.5cm}
\def\f{5cm} 
\def\g{6cm} \def\G{6.5cm}
 
\def\z{8cm}

\def\sizePoint{3pt}

\newcommand{\point}[2]{\fill (canvas cs:x=#1,y=#2) circle (\sizePoint); }
\def\styleGrille{densely dotted}
\title{Catalan and Schr\"oder permutations sortable by two restricted stacks}
\author[1]{Jean-Luc Baril}
\author[2]{Giulio Cerbai\thanks{G.C. is member of the INdAM Research group GNCS; he is partially supported by INdAM-GNCS 2020 project ``Combinatoria delle permutazioni, delle parole e dei grafi: algoritmi e applicazioni''.}}
\author[1]{Carine Khalil}
\author[1]{Vincent Vajnovszki}
\affil[1]{LIB, Universit\'e de Bourgogne Franche-Comt\'e,
B.P. 47 870, 21078 Dijon Cedex France {\tt \{barjl,carine.khalil,vvajnov\}@u-bourgogne.fr}
} 
\affil[2]{Dipartimento di Matematica e Informatica ``U. Dini'', University of Firenze, Firenze, Italy, {\tt giulio.cerbai@unifi.it}}
\date{}

\begin{document}
\maketitle

\begin{abstract}
Pattern avoiding machines were introduced recently by Claesson, Cerbai and Ferrari 
as a particular case of the two-stacks in series sorting device.
They consist of two restricted stacks in series, ruled by a right-greedy procedure and the stacks avoid some specified  patterns.
Some of the obtained results have been further generalized to Cayley permutations by Cerbai, specialized to particular patterns by Defant and Zheng, or considered in the context of functions over the symmetric group by Berlow.
In this work we study pattern avoiding machines where the first stack avoids a pair of patterns of length 3 and investigate those pairs for which sortable permutations are counted by the (binomial transform of the) Catalan numbers and the Schr\"oder numbers.
\end{abstract}

\section{Introduction}

Pattern avoiding machines were recently introduced in~\cite{CCF20} in attempt to gain a better understanding of sortable permutations using stacks in series. They consist of two restricted stacks in series, equipped with a right-greedy procedure, where the first stack avoids a fixed pattern, reading the elements from top to bottom; and the second stack avoids the pattern $21$ (which is a necessary condition for the machine to sort permutations).
The authors of~\cite{CCF20} provide a characterization of the avoided patterns for which sortable permutations 
do not form a class, and they show that those patterns are enumerated by the Catalan numbers. For specific patterns, such as $123$ and the decreasing pattern of any length, a geometrical description of sortable permutations is also obtained. The pattern $132$ has been solved later in~\cite{CCFS20}. Some of these results have been further generalized to Cayley permutations in~\cite{C_Cayley}. 
More recently, Berlow~\cite{Berlow20} explores a single stack version of pattern avoiding machines, where the stack avoids a set of patterns and the sorting process is regarded as a function. Analogous machines, but based on the notion of consecutive patterns, have been introduced and discussed in~\cite{Def}.

In this work we study a variant of pattern-avoiding machines where the first stack avoids 
$(\sigma,\tau)$, a pair of patterns  of length three. Following~\cite{CCF20}, we call it $(\sigma,\tau)$-machine.
More specifically, we restrict ourselves to those pairs of patterns for which sortable permutations are counted by either the Catalan numbers or two of their close relatives: the binomial transform of Catalan numbers and the Schr\"oder numbers. For the pair $(132,231)$ we show that sortable permutations are those avoiding $1324$ and $2314$, a set whose enumeration is given by the large Schr\"oder numbers. 
Under certain conditions on the avoided patterns, the output of the first stack is bijectively related to its input (see~\cite{Berlow20,C_Cayley}): it follows that for three pairs of patterns, namely $(123,213)$, $(132,312)$ and $(231,321)$, sortable permutations are counted by the Catalan numbers. This result was proved independently in~\cite{BKV_PP20,Berlow20}.
For the pair $(123,132)$, we prove that sortable permutations are those avoiding the patterns $2314$, $3214$, $4213$ and the generalized pattern $[24\bar{1}3$. We prove that sortable permutations are enumerated by the Catalan numbers by showing that the distribution of the first element is given by the well-known Catalan triangle. 
Finally, we show that for the pair $(123, 312)$ the corresponding counting sequence is the binomial transform of Catalan numbers.

This paper is the extended version of the conference presentation
\cite{BKV_PP20}.

\section{Notations and some preliminary results}\label{section_notations}

We start by recalling some classical definitions about pattern avoidance on permutations (see~\cite{Kit} for a more detailed introduction). Denote by $\mathfrak{S}_n$ the set of permutations of length $n$ and let $\mathfrak S=\cup_{n\geq 0}\mathfrak S_n$. Given two permutations $\sigma$ of length $k$ and $\pi=\pi_1\cdots\pi_n$, we say that $\pi$ {\it contains} the pattern $\sigma$ if $\pi$ contains a subsequence $\pi_{i_1}\cdots\pi_{i_k}$, with $i_1<i_2<\cdots<i_k$, which is order isomorphic to $\sigma$. In this case, we say that $\pi_{i_1}\cdots\pi_{i_k}$ is an {\it occurrence} of the pattern $\sigma$ in $\pi$. Otherwise, we say that $\pi$ {\it avoids}~$\sigma$.

We say that $\pi$ contains an occurrence of the (generalized) pattern $[\sigma$ if $\pi$ contains an occurrence of $\sigma$ that involves the first element $\pi_1$ of $\pi$. For instance, an occurrence of $[12$ in $\pi$ corresponds to a pair of elements $\pi_1\pi_i$, with $i>1$ and $\pi_i>\pi_1$. A {\it barred pattern} $\tilde{\sigma}$ is a pattern where some entries are barred. Let $\sigma$ be the classical pattern obtained by removing all the bars from $\tilde{\sigma}$. Let $\tau$ be the pattern which is order isomorphic to the non-barred entries of $\tilde{\sigma}$ (i.e. obtained from $\tilde{\sigma}$ by removing all the barred entries and suitably rescaling the remaining elements). A permutation $\pi$ avoids $\tilde{\sigma}$ if each occurrence of $\tau$ in $\pi$ can be extended to an occurrence of $\sigma$. For instance, a permutation $\pi$ avoids the pattern $[24\bar{1}3$ if for any subsequence $\pi_1 \pi_i \pi_j$, with $1<i<j$ and $\pi_1<\pi_j<\pi_i$, there is an index $t$, $i<t<j$, such that $\pi_1\pi_i\pi_t\pi_j$ is an occurrence of $2413$.

Given a set of (generalized) patterns $T$, denote by $\Av_n(T)$ the set of permutations in $\mathfrak S_n$ avoiding each pattern in $T$. Similarly, let $\Av(T)=\cup_{n\geq 0}\Av_n(T)$. If $T=\{ \sigma \}$ is a singleton, we write $\Av_n(\sigma)$ and $\Av(\sigma)$. In his celebrated book~\cite{Knuth68}, Knuth gave the following characterization of stack sortable permutations, which is often considered the starting point of stack sorting and permutation patterns disciplines. 

\begin{prop}[\cite{Knuth68}]
A permutation $\pi$ is sortable using a classical stack (that is, a $21$-avoiding stack) if
and only if $\pi$ avoids the pattern $231$.
\label{KnuthPr}
\end{prop}

Let $T$ be a set of patterns. A $T$-stack is a stack that is not allowed to contain an occurrence of any pattern in $T$, reading its elements from top to bottom. Given a permutation $\pi$, denote by $\out^T(\pi)$ the permutation obtained after passing $\pi$ through the $T$-avoiding stack by applying a greedy procedure, i.e. by always pushing the next element of the input, unless it creates an occurrence of a forbidden pattern inside the stack. Denote by $\Sort_n(T)$ the set of length $n$ permutations that are sortable by the $T$-machine, that is, by passing $\pi$ through the $T$-avoiding stack and then through the $21$-avoiding stack. Permutations in $\Sort_n(T)$ are called {\it $T$-sortable}, and $\Sort(T)$ is the set of $T$-sortable permutations of any length.
As a consequence of Proposition~\ref{KnuthPr}, $\Sort(T)$ consists precisely of those permutations $\pi$ for which $\out^{T}(\pi)$ avoids $231$. To ease notations, if $T$ is either a singleton $T=\lbrace \sigma\rbrace$ or a pair of patterns $T=\lbrace \sigma,\tau \rbrace$, we will omit the curly brackets from the above notations. For instance, we will write $\Sort(\sigma,\tau)$ instead of $\Sort(\lbrace \sigma,\tau\rbrace)$.

The authors of~\cite{CCF20} showed that if $\pi$ is a $12$-sortable permutation of length $n$, then $\out^{12}(\pi)=n(n-1)\cdots 1$.
Moreover, by Proposition~\ref{KnuthPr} and applying the complement operation on the processed permutation, we have that $\Sort(12)=\mathfrak{S}(213)$. 
In order to refer to this result later, we state it below in a slightly more general form. A \textit{partial permutation} of $n$ is an injection $\pi:\{1,2,\dots,k\}\to\{1, 2,\dots, n\}$, for some $0\leq k \leq n$, and the integer $k$ is said to be the length of $\pi$. We let a $12$-stack act on a partial permutation $\pi$ of $n$ in the natural way by identifying $\pi$ with the list of its images.

\begin{prop}
If $\pi$ is a partial permutation of $n$ which is $12$-sortable, then $\out^{12}(\pi)$
is the decreasing rearrangement of the symbols of $\pi$. Moreover, $\pi$ is $12$-sortable if and only if it avoids $213$.
\label{arrangements}
\end{prop}

An entry $\pi_i$ of a permutation $\pi$ is a \textit{left-to-right minimum} if $\pi_i < \pi_j$, for each $j<i$. The \textit{left-to-right minima decomposition} (briefly \textit{ltr-min decomposition}) of $\pi$ is $\pi=m_1 B_1 m_2 B_2 \cdots m_t B_t$, where $m_1>m_2>\cdots >m_t$ are the ltr-minima of $\pi$ and the \textit{block} $B_i$ contains the elements of $\pi$ between $m_{i}$ and $m_{i+1}$, for $i=1,\dots,t-1$. The last block $B_t$ contains the elements that follow $m_t$ in $\pi$. Note that $m_t=1$. The notion of \textit{left-to-right maximum} of a permutation $\pi$ is defined similarly. The \textit{ltr-max decomposition} of $\pi$ is $\pi=M_1 B_1 M_2 B_2 \cdots M_t B_t$, where $M_1<M_2<\cdots <M_t$ are the ltr-maxima of $\pi$. In this case $M_t=n$, where $n$ is the length of $\pi$.

Finally, the sequence $(c_n)_{n\geq 0}$, pervasive in this paper, is the sequence of Catalan numbers $c_n=\frac{1}{n+1}{2n\choose n}$ ({\tt A000108} in~\cite{Sloane}).

\section{Pair $(132,231)$}
This section is devoted to the analysis of the $(123,231)$-machine.

\begin{theorem}\label{theorem_ltr-min_pairs_of_pat_132}
Consider the $(132,\sigma)$-machine, where $\sigma=\sigma_1 \cdots \sigma_{k-1} \sigma_k \in \mathfrak{S}_k$, with $k \ge 3$ and
$\sigma_{k-1}>\sigma_k$. Given a permutation $\pi$ of length $n$, let $m_1 B_1 \cdots m_t B_t=\pi$ be its ltr-min decomposition. Then:
\begin{enumerate}
\item Everytime a ltr-minimum $m_i$ is pushed into the $(132,\sigma)$-stack, the $(132,\sigma)$-stack contains the elements $m_{i-1},\dots,m_2,m_1$, reading from top to bottom. Moreover, we have
$$
\out^{132,\sigma}(\pi)=\tilde{B}_1\cdots\tilde{B}_tm_t\cdots m_1,
$$
where $\tilde{B}_i$ is a rearrangement of $B_i$.
\item If $\pi$ is $(132,\sigma)$-sortable, then $\tilde{B}_i$ is decreasing for each $i$. Moreover, for each $i\le t-1$, we have $B_i>B_{i+1}$ (i.e. $x>y$ for each $x\in B_i, y\in B_{i+1}$).
\end{enumerate}
\end{theorem}

\proof
\comm{
\begin{enumerate}
\item Let us consider the evolution of the $(132,\sigma)$-stack on input $\pi$. Note that, since $k \ge 3$, the element $m_1$ remains at the bottom of the $(132,\sigma)$-stack until the end of the process. Now, if $B_1$ is not empty then for each $x \in B_1$, the elements $m_2 x m_1$ form an occurrence of $132$. Therefore the block $B_1$ is extracted before $m_2$ enters the $(132,\sigma)$-stack. After $m_2$ is pushed, the $(132,\sigma)$-stack contains $m_2 m_1$, reading from top to bottom. 
Since $m_2 < m_1$, but $\sigma_{k-1} > \sigma_k$ by hypothesis, $m_2$ cannot play the role of either $\sigma_{k-1}$ in an occurrence of $\sigma$ or of $3$ in an occurrence of $132$. Thus $m_2$ remains at the bottom of the $(132,\sigma)$-stack until the end of the sorting procedure. The thesis follows by iterating the same argument on each block $B_i$, for $i \ge 2$.
\item Suppose that $\pi$ is $(132,\sigma)$-sortable. Assume, for a contradiction, that $\tilde{B}_i$ is not decreasing, for some $i$. Then there are two consecutive elements $x<y$ in $\tilde{B}_i$. Therefore, by what proved above, $\out^{132,\sigma}(\pi)$ contains an occurrence $xym_t$ of $231$, which is impossible due to Proposition~\ref{KnuthPr}. Finally, suppose that $x<y$, for $x\in B_i$ and $y\in B_{i+1}$. Then $xym_t$ is an occurrence of $231$ in $\out^{132,\sigma}(\pi)$, a contradiction.
\end{enumerate}
}
\endproof

Theorem~\ref{theorem_ltr-min_pairs_of_pat_132} and Proposition~\ref{arrangements} guarantee that if $\pi=m_1 B_1 \cdots m_t B_t$ is the the ltr-min decomposition of a $(132,231)$-sortable permutation $\pi$, then (with the notation above) $\tilde{B}_i = \out^{12} (B_i)$, for each $i$. However, this is true even when the sortability requirement is relaxed.

\begin{lemma}\label{132-231_equivalence}
Let $\pi=m_1 B_1 \cdots m_t B_t$ be the ltr-min decomposition of a permutation $\pi$. Write $\out^{132,231}(\pi)= \tilde{B}_1 \cdots \tilde{B}_t m_t \cdots m_1$ as in Theorem~\ref{theorem_ltr-min_pairs_of_pat_132}. Then $\tilde{B}_i = \out^{12} (B_i)$, for each $i$.
\end{lemma}
\proof \comm{
Consider the instant immediately after $m_i$ is pushed into the $(132,231)$-stack and the non-empty block 
$B_i$ has to be processed, for some $i$. By Theorem~\ref{theorem_ltr-min_pairs_of_pat_132}, at this point the $(132,231)$-stack contains $m_i, m_{i-1},\dots,m_1$, reading from top to bottom. We want to show that the behavior of the $(132,231)$-stack on $B_i$ is equivalent to the behavior of an empty $12$-stack on input $B_i$. We prove that the $(132,231)$-stack performs the pop operation of some $x \in B_i$ if and only if the $12$-stack does the same. If either the next element of the input is $m_{i+1}$ or $x$ is the last element of $\pi$ to be processed, then both the $(132,231)$-stack and the $12$-stack perform a pop operation, as desired. Otherwise, suppose the next element of the input is $y$, for some $y$ in the same block $B_i$, and the $(132,231)$-stack pops the element $x \in B_i$. This means that the $(132,231)$-stack contains two elements $z,w$, with $z$ above $w$, such that $yzw$ is an occurrence of either $132$ or $231$. Note that, since $z>w$, $z$ is not a ltr-minimum. Therefore $yz$ is an occurrence of $12$ and the $12$-stack performs a pop operation, as desired. Conversely, suppose that the $12$-stack pops the element $x$, with $y \in B_i$ the next element of the input. This implies that the $12$-stack contains an element $z$ such that $z>y$. Therefore $y z m_i$ is an occurrence of $231$ and the $(132,231)$-stack performs a pop operation, as desired.
}
\endproof

\begin{corollary}
\label{three_cases}
Let $\pi=m_1 B_1 \cdots m_t B_t$ be the ltr-min decomposition of a permutation $\pi$. Then the following are equivalent.
\begin{enumerate}
\item $B_i$ avoids $213$ and $B_i > B_{i+1}$, for each $i$.
\item $\pi$ is $(132,231)$-sortable.
\item $\pi \in \Av(1324,2314)$.
\end{enumerate}
\end{corollary}
\proof\comm{
Combining the first point in Theorem~\ref{theorem_ltr-min_pairs_of_pat_132} and Lemma~\ref{132-231_equivalence} we have:

$$\out^{132,231}(\pi) = \out^{12}(B_1) \cdots \out^{12}(B_t) m_t \cdots m_1.$$

We will use this decomposition of $\out^{132,231}(\pi)$ throughout the rest of the proof.

\medskip
\noindent
$\left[ 1 \Rightarrow 2 \right]$ 
Suppose, for a contradiction, that $\out^{132,231}(\pi)$ contains an occurrence $bca$ of $231$. Note that, since $c>a$, while $m_t<\cdots<m_1$, $c$ is not a ltr-minimum of $\pi$ (and thus neither $b$ is). Now, if $b$ and $c$ are in the same block $B_j$, then $\out^{12}(B_j)$ is not decreasing. Thus, by Proposition 
\ref{arrangements}, $B_j$ contains $213$, which is a contradiction. Otherwise, if $b \in B_j$ and $c \in B_k$, with $j<k$, then we have a contradiction with the hypothesis $B_i > B_{i+1}$ for each $i$.

\noindent
$\left[ 2 \Rightarrow 3 \right]$ Suppose, for a contradiction, that $\pi \notin \Av(1324,2314)$. First, suppose that $\pi$ contains an occurrence $acbd$ of $1324$. Observe that $b,c,d$ are not ltr-minima of $\pi$. Now, if $b$ and $d$ are in the same block $B_j$ of $\pi$, for some $j$, then $B_j$ contains an occurrence $cbd$ of $213$. Therefore $\out^{12}(B_j)$ contains an occurrence of $231$ due to
Proposition~\ref{arrangements}, which contradicts the hypothesis. Otherwise, if $b \in B_j$ and $d \in B_k$, for some $j<k$, then $\out^{132,231}(\pi)$ contains an occurrence $bd m_k$ of $231$, again a contradiction. The pattern $2314$ can be addressed analogously, so we leave it to the reader.

\noindent
$\left[ 3 \Rightarrow 1 \right]$ Let $\pi\in\Av(1324,2314)$. If $B_i$ contains an occurrence $bac$ of $213$, then $\pi$ contains an occurrence $m_i bac$ of $1324$, which is impossible. Otherwise, if $\pi$ contains two elements $x \in B_j$, $y \in B_k$, with $x<y$ and $j<k$, then $m_j x m_k y$ is an occurrence of $2314$, contradicting the hypothesis.
}
\endproof

The enumeration of $\Av(1324,2314)$ (or a symmetry of these patterns) can be found for instance in~\cite{Barc,Wes1}. Note that in~\cite{Ban}, the authors provide a constructive bijection between these permutations and Schr\"oder paths.

\begin{corollary} Permutations of length $n$ in  $Sort(132,231)$ are enumerated by the large Schr\"oder numbers (sequence {\tt A006318} in~\cite{Sloane}).
\end{corollary}


\section{The $(\sigma,\hat{\sigma})$-machine}

For a permutation $\sigma$ of length two or more, denote by $\hat\sigma$ the permutation obtained from $\sigma$ by interchanging its first two entries. Let us regard a $(\sigma,\tau)$-stack as an operator $\out^{\sigma,\tau}: \mathfrak{S}\to\mathfrak{S}$. By conveniently modifying the proof of Corollary~4.5 in~\cite{C_Cayley} (stated in the context of Cayley permutations), we have that $\out^{\sigma,\tau}$ is a length preserving bijection on $\mathfrak S$ if and only if $\tau=\hat\sigma$. More generally, Berlow~\cite{Berlow20} showed that for a set $T$ of patterns, $\out^T$ is a length preserving bijection on $\mathfrak{S}$ if and only if
$T$ is closed under the~$\hat{ \phantom{u}}$~operator.
In order for the paper to be self-contained, we shall give the following result, which is easier to prove (although weaker): $\out^{\sigma,\hat\sigma}$ is a bijection for any pattern $\sigma$. An immediate consequence will be Theorem~\ref{th_sigma_hat_sigma} below.
\medskip

Let $N^*$ be the set of finite length integer sequences. The action of the $(\sigma,\tau)$-stack on input $\pi$ can be naturally represented as a sequence of triples $(r;s;t)\in \left(N^*\right)^3$, where $r$ is the current content of the output, $s$ is the current content of the $(\sigma,\tau)$-stack (read from top to bottom) and $t$ is the current content of the input. The triple $(r;s;t)$ is said to be a {\it state of passing} of $\pi$ through the $(\sigma,\tau)$-stack.
Clearly, $r$ is a prefix of $\out^{\sigma,\tau}(\pi)$, $t$ is a suffix of $\pi$, the initial state is $(\lambda;\lambda;\pi)$ and the final one is $(\out^{\sigma,\tau}(\pi);\lambda;\lambda)$, where $\lambda$ is the empty sequence.
Moreover a non-final state $(p_1p_2\cdots p_a;s_1s_2\cdots s_b;t_1t_2\cdots t_c)$ is followed by either the state 
$$(p_1p_2\cdots p_as_1;s_2\cdots s_b;t_1t_2\cdots t_c),$$
if a pop operation is performed next, or
$$(p_1p_2\cdots p_a;t_1s_1s_2\cdots s_b;t_2\cdots t_c),$$
if a push operation is performed next.

For $p=p_1\cdots p_n\in\mathbb N^n$, we denote by $p^\mathrm r$ the {\it reverse} of $p$, that is $p^\mathrm r=p_n\cdots p_1$. 
We wish to show that the behavior of the $(\sigma,\hat{\sigma})$-stack on $\pi$ is strictly related to its behavior on $\left(\out^{\sigma,\hat\sigma}(\pi)\right)^{\mathrm r}$. More precisely, if $o_1\cdots o_{2n}$ is the sequence of push/pop operations performed when $\pi$ is passed through a $(\sigma,\hat\sigma)$-stack, then $o'_{2n}\cdots o'_1$ is the sequence of push/pop operations performed when $\left(\out^{\sigma,\hat\sigma}(\pi)\right)^{\mathrm r}$ is passed through the $(\sigma,\hat\sigma)$-stack, where $o'_i$ is a push (resp. pop) operation if $o_i$ is a pop (resp. push) operation.
This can be equivalently expressed by saying that the state $(p;s;t)$ is followed by $(u;v;w)$ if and only if the state $(w^{\mathrm r};v;u^{\mathrm r})$ is followed by $(t^{\mathrm r};s;p^{\mathrm r})$.

\begin{lemma}
Consider the action of the $(\sigma,\hat\sigma)$-stack. Let $p,s,t\in\mathbb N^*$ and $x\in\mathbb N$.
\begin{enumerate}
\item 
If the state $(p,xs,t)$ is followed by the state $(px,s,t)$ (and thus a pop operation is performed) then the state
$(t^{\mathrm r},s,xp^{\mathrm r})$ is followed by the state $(t^{\mathrm r},x s, p^{\mathrm r})$ (and thus a push operation is performed).
\item 
If the state $(p,s, x t)$ is followed by the state $(p,x s, t)$ (and thus a push operation is performed), then the state 
$(t^{\mathrm r},x s,p^{\mathrm r})$ is followed by the state $(t^{\mathrm r} x, s,p^{\mathrm r})$ (and thus a pop operation is performed).
\end{enumerate}
\end{lemma}
\proof\comm{
\begin{enumerate}
\item Since $xs$ is the content of the $(\sigma,\hat{\sigma})$-stack in the state $(p,xs,t)$, we have that $xs$ avoids $\sigma$ and $\hat\sigma$. Thus a push operation is performed if $s$ is the content of the $(\sigma,\hat{\sigma})$-stack and $x$ is the next element of the input.

\item If $p$ is empty, the statement holds. Otherwise, let $p=p_1\cdots p_a$ and
$s=s_1\cdots s_b$. Observe that $p_a$ is the last element that has been extracted from the $(\sigma,\hat{\sigma})$-stack before $x$ enters. Therefore, when $p_a$ is extracted, $p_a$ plays the role of either $\sigma_2$ in an occurrence of $\sigma$ or of $\hat{\sigma}_2$ in an occurrence of $\hat{\sigma}$. More precisely, one of the following four cases hold. We show the details for the first case only, the others being similar. Let $z$ be the length of $\sigma$.
\begin{itemize}
\item $s_{\ell}p_as_{i_3}\cdots s_{i_z}$ is an occurrence of $\sigma$, for some $\ell\ge 1$ and $\ell<i_3<\cdots<i_z$. Then $p_as_{\ell}s_{i_3}\cdots s_{i_z}$ is an occurrence of $\hat{\sigma}$ and therefore a pop operation is performed when $p_a$ is the next element of the input and $xs$ is the content of the $(\sigma,\hat{\sigma})$-stack, as desired.
\item $s_{\ell}p_as_{i_3}\cdots s_{i_z}$ is an occurrence of $\hat{\sigma}$, for some $\ell\ge 1$ and $\ell<i_3<\cdots<i_z$.
\item $xp_as_{i_3}\cdots s_{i_z}$ is an occurrence of $\sigma$, for some $i_3<\cdots<i_z$.
\item $xp_as_{i_3}\cdots s_{i_z}$ is an occurrence of $\hat{\sigma}$, for some $i_3<\cdots<i_z$.
\end{itemize}

\end{enumerate}
}
\endproof

A straightforward consequence of the previous lemma is that the map $\left(\out^{\sigma,\hat\sigma}\right)^{\mathrm r}:\mathfrak S\to \mathfrak S$ is its own inverse, and thus a bijection. More specifically, for any permutation $\pi$, we have $\left(\out^{\sigma,\hat\sigma}\right)^{-1}(\pi)=\left(\out^{\sigma,\hat\sigma}(\pi^{\mathrm r})\right)^{\mathrm r}$. Since $\pi$ is $(\sigma,\hat{\sigma})$-sortable if and only if $\out^{\sigma,\hat{\sigma}}(\pi)$ avoids $231$ (and the reverse map is bijective), we have that $\Sort(\sigma,\hat \sigma)$ is in bijection with $\Av(231)$. The next theorem follows.

\begin{theorem}
For any pattern $\sigma$, $\out_n^{\sigma,\hat{\sigma}}$ is a bijection on $\mathfrak S_n$. Moreover, we have
$$|\Sort_n(123,213)|=|\Sort_n(132,312)|=|\Sort_n(231,321)|=c_n,$$
the $n${\em th} Catalan number.
\label{th_sigma_hat_sigma}
\end{theorem}


\section{Pair $(123, 132)$}

We characterize $\Sort(123,132)$ in terms of pattern avoidance. Then we show that $(123,132)$-sortable permutations are enumerated by the Catalan numbers by exhibiting a link with the very well studied Catalan triangle.

\begin{theorem} 
A permutation $\pi$ is $(123,132)$-sortable if and only if $\pi$ avoids $2314$,$3214$, $4213$ and $[24\bar{1}3$.
\label{Th1}
\end{theorem}

\proof
\comm{ 
Suppose that $\pi$ is $(123, 132)$-sortable. For a contradiction, suppose that $\pi$ contains 
$\tau\in\{2314,3214,4213\}$.  Pick an occurrence $\pi_i\pi_j\pi_k\pi_\ell$ of $\tau$, with $i<j<k<\ell$, where $\ell$ is chosen  minimal, and $k$, $j$, and $i$ are chosen maximal, in this  order.
\medskip

\noindent
If $\tau=2314$, due to our choice of $i,j,k,\ell$, we have $\pi_i<\pi_u<\pi_j$, for $k<u<\ell$.
Now, when $\pi_k$ is pushed in the $(123,132)$-stack, at least one of $\pi_i$ and $\pi_j$ has already been extracted: otherwise the $(123,132)$-stack would contain an occurrence of $132$, which is forbidden. For each $u$, $k+1\leq u\leq \ell$, we have 
$\pi_k<\pi_u$ and when $\pi_u$ is pushed in the $(123,132)$-stack, $\pi_k$ is still in the $(123,132)$-stack. Indeed, 
$\pi_u\pi_x\pi_y$ cannot be an occurrence of $123$ nor of $132$ with $\pi_x$ above $\pi_y$, both in the tail of the $(123,132)$-stack beginning by $\pi_k$.
If $\pi_i$ (resp. $\pi_j$) is extracted before $\pi_k$ enters in the $(123,132)$-stack, then 
$\pi_i\pi_{k+1}\pi_k$ (resp. $\pi_j\pi_\ell\pi_k$) creates a pattern $231$ in $\out^{123,132}(\pi)$, a contradiction.
\medskip

\noindent
If $\tau=3214$, due to our choice of $i,j,k,\ell$, we have $\pi_i>\pi_u>\pi_j$, for $k<u<\ell$; and $\pi_u<\pi_{u+1}$, for $k\leq u<\ell$. As above, when $\pi_k$ is pushed in the $(123,132)$-stack, at least one of $\pi_i$ and $\pi_j$ has already been extracted: otherwise the $(123,132)$-stack would contain an occurrence of $123$. Since $\pi_{k+1}>\pi_k$, the next step pushes $\pi_{k+1}$ in the $(123,132)$-stack. 
($i$) Assume that $\pi_i$ and $\pi_j$ both had left the $(123,132)$-stack. 
Then $\pi_k$ is just below $\pi_{k+1}$ in the $(123,132)$-stack, and $\pi_k<\pi_j<\pi_{k+1}$.
This implies that $\pi_j\pi_{k+1}\pi_k$ is an occurrence of $231$ in
$\out^{123,132}(\pi)$, a contradiction. 
($ii$) Assume that $\pi_i$ is still in the $(123,132)$-stack and $\pi_j$ had left this stack.
Again, $\pi_j\pi_{k+1}\pi_k$ is an occurrence of $231$ in $\out^{123,132}(\pi)$, a contradiction.
($iii$) Assume that $\pi_j$ is still in the $(123,132)$-stack and $\pi_i$ had left this stack. Since $\pi_u<\pi_{u+1}$ for $k\leq u\leq \ell-1$, the next steps of the process push successively all entries $\pi_{k+1}, \ldots, \pi_{\ell}$ in the $(123,132)$-stack.
As above, $\pi_i\pi_\ell\pi_k$ is an occurrence of $231$ in $\out^{123,132}(\pi)$, again a contradiction.
\medskip

\noindent
The case $\tau=4213$ can be treated similarly. 

\medskip

Finally, suppose that $\pi$ contains $[24\bar{1}3$. Equivalently, there are two indices $i<j$ such that $\pi_1\pi_i\pi_j$ is an occurrence of $132$ and $\pi_k>\pi_1$ for each $i<k<j$. Observe that, by choosing $j$ minimal and $i$ maximal (in this order), we can assume $j=i+1$. Now, if $\pi_i$ is still in the $(123,132)$-stack when $\pi_{i+1}$ enters, then $\out^{123,132}(\pi)$ contains an occurrence $\pi_{i+1}\pi_i\pi_1$ of $231$, which is impossible due to the sortability of $\pi$. Therefore $\pi_i$ is extracted before $\pi_{i+1}$ enters. This means that there are two elements $\pi_u,\pi_v$ in the $(123,132)$-stack, with $u<v$ (and thus $\pi_v$ above $\pi_u$), such that $\pi_{i+1}\pi_v\pi_u$ is an occurrence of either $123$ or $132$. Choose $u,v$ minimal amongst those indices\footnote{In other words, pick the deepest such elements $\pi_u,\pi_v$ in the $(123,132)$-stack}, so that $\pi_u$ is still in the $(123,132)$-stack when $\pi_{i+1}$ enters. Notice that $\pi_{i+1}<\pi_u$ (and thus $u\neq 1$). Moreover, it must be $\pi_u<\pi_i$, otherwise $\pi_i\pi_u\pi_1$ would be an occurrence of $231$ in the $(123,132)$-stack, which is forbidden. But then $\pi_{i+1}\pi_u\pi_1$ is an occurrence of $231$ in $\out^{123,132}(\pi)$, a contradiction.

Conversely, suppose that $\pi$ is not $(123,132)$-sortable. We shall prove that $\pi$ contains at least one of the patterns $3214$, $2314$, $4213$ or $[24\bar{1}3$. By hypothesis $\out^{123,132}(\pi)$ contains an occurrence $bca$ of $231$. Let $b=\pi_j$ and $c=\pi_k$, for some indices $j,k$. We distinguish two cases, according whether $j<k$ or $j>k$.
\begin{itemize}
\item Suppose that $j<k$ and thus $\pi_j$ is extracted from the $(123,132)$-stack before $\pi_k$ enters. Then there are two elements $\pi_u,\pi_v$ in the $(123,132)$-stack, with $u<v$ (and thus $\pi_v$ above $\pi_u$), such that $\pi_z\pi_v\pi_u$ is an occurrence of either $123$ or $132$, where $\pi_z$ is the next element of the input. Notice that $\pi_j\ge\min\{ \pi_u,\pi_v\}$, since otherwise $\pi_j\pi_v\pi_u$ would be an occurrence of either $123$ or $132$ in the $(123,132)$-stack, which is impossible. Thus $\pi_k>\pi_j>\pi_z$. If $\pi_z\pi_v\pi_u$ is an occurrence of $123$, then $\pi_u\pi_v\pi_z\pi_k$ is an occurrence of either $4213$, if $\pi_u>\pi_k$, or $3214$, if $\pi_u<\pi_k$. Finally, if $\pi_z\pi_v\pi_u$ is an occurrence of $132$, then $\pi_u\pi_j\pi_z\pi_k$ is an occurrence of $2314$.
\item Suppose instead that $j>k$ and thus $\pi_k$ is still in the $(123,132)$-stack when $\pi_j$ enters. Observe that $k\neq 1$, since $\pi_1$ is the last element of $\out^{123,132}(\pi)$. Therefore, when $\pi_j$ enters the $(123,132)$-stack, the $(123,132)$-stack contains the elements $\pi_j\pi_k\pi_1$, reading from top to bottom. Notice that $\pi_j>\pi_1$, otherwise $\pi_j\pi_k\pi_1$ would realize an occurrence of the forbidden $132$ inside the $(123,132)$-stack. Moreover, for each entry $\pi_t$, with $k<t<j$, we have $\pi_t>\pi_1$. Otherwise $\pi_t\pi_k\pi_1$ would be an occurrence of $132$ and $\pi_k$ would be extracted before $\pi_j$, which is impossible due to our assumptions. Thus $\pi_1\pi_k\pi_j$ is an occurrence of $[24\bar{1}3$. This completes the proof.
\end{itemize}
}
\endproof

\begin{prop}
The distribution of the first element in $\Sort(123,132)$
is given by the Catalan triangle (sequence {\tt A009766} in~\cite{Sloane}). 
\end{prop}
\proof\comm{
Let $A_n(k)$ be the set of $(123,132)$-sortable permutations of length $n$ and starting with $k$. Let $A_n^1(k)$ be the subset of $A_n(k)$ consisting of those permutations $\pi=\pi_1\pi_2\ldots \pi_n$ where any occurrence $\pi_1\pi_i\pi_\ell$ of $[231$ with $\pi_\ell=\pi_1-1$ can be extended into an occurrence  $\pi_1\pi_i\pi_j\pi_\ell$ of $[3412$. Set $A_n^2(k)=A_n(k)\backslash A_n^1(k)$ and let $k\ge 2$. We shall provide bijections $\alpha:A_n^1(k)\to A_n(k-1)$ and $\beta:A_n^2(k)\to A_{n-1}(k)$.

Define $\alpha:A_n^1(k)\to A_n(k-1)$ by $\alpha(\pi)=\pi'$, where $\pi'$ is obtained from $\pi$ by swapping the two entries $\pi_1$ and $\pi_1-1$ in $\pi$. Since $\pi\in A_n^1(k)$, it is easy to check that $\pi'$ avoids $[24\bar{1}3$. In addition, swapping $\pi_1$ and $\pi_1-1$ does not affect the avoidance of the three patterns $3214$, $2314$, $4213$, which implies (see Theorem~\ref{Th1}) that $\alpha(\pi)\in A_n(k-1)$.
%

Next define $\beta:A_n^2(k)\to A_{n-1}(k)$ by $\beta(\pi)=\pi''$, where $\pi''$ is obtained from $\pi$ by deleting the entry $\pi_{\ell}$ immediately before $k-1$ and by decreasing by one all entries of $\pi$ greater than $\pi_{\ell}$. Notice that $\beta(\pi)\in A_{n-1}(k)$. Let us now sketch the proof that $\beta$ is bijective, leaving some technical details to the reader. We shall explicitly define the inverse map of $\beta$. Given $\pi\in A_{n-1}(k)$, choose an integer $\ell$ as follows:

\begin{itemize}
\item $\ell$ is the minimal entry $l=\pi_u>\pi_1$, with $1<u<i$, such that there is an index $v$ with $\pi_v<\pi_i$ and $u<v<i$, if such entry exists.
\item Otherwise, set $\ell=n$.
\end{itemize}

The preimage $\pi$ is obtained by inserting $\ell$ immediately before $\pi_i=k-1$ and then increasing by one all the entries $\pi_j$ of $\pi$ with $\pi_j\ge\ell$.

Finally, setting $a_n^k=|A_n(k)|$, we have that $a_n^k=a_n^{k-1}+a_{n-1}^k$, for $2\leq k\leq n$. Since $A_n(1)=\{123\cdots n\}$ and $A_n(n)$ is the set of length $n$ permutations avoiding $213$ and starting with $n$, the initial conditions are given by $a_n^1=1$ and $a_n^n=c_{n-1}$, where $c_n$ is the $n$th Catalan number. Therefore, $a_n^k$ generates the well-known Catalan triangle (see Table~\ref{t2} and~\cite{Cai,Des,Sha}).
}
\endproof

\begin{table}[h]
\centering
\begin{tabular}{c|*{8}{m{0.7cm}}}
   $k \backslash n$  & 1&2 &3 &4 &5 &6 &7&8\\
     \hline
                    1 & 1&1 &1 &1 &1 &1 &1&1\\

                    2 & &1&2 &3 &4 &5 &6 &7\\

                    3 &  & &2 &5 &9 &14 &20 &27\\

                    4 & & &&5 &14 &28 &48&75\\

                   5 & & & & &14 &42 &90&165\\

                     6 & & & & & &42 &132&297\\
                      $\ldots$ & & & & & &&$\ldots$ &$\ldots$\\
     \hline
      $\sum$ & 1&2 &5 &14 &42 &132 &429&1430\\
\end{tabular}
\vskip0.3cm
\caption{The Catalan triangle $a_n^k=|A_n(k)|$, with $1\leq n\leq 8$ and $1\leq k \leq 6$.
}
\label{t2}
\end{table}

\begin{corollary}
Permutations of length $n$ in  $\Sort(123,132)$ are enumerated by the Catalan numbers.
\end{corollary}
\proof
With the previous notations we have $|\Sort_n(123,132)|=\sum_{k=1}^na_n^k=c_n$, the $n$th Catalan number (see again~\cite{Cai,Des,Sha}).
\endproof

\section{Pair $(123,312)$}

We start by giving a ltr-max counterpart of Theorem~\ref{theorem_ltr-min_pairs_of_pat_132}.

\begin{theorem}\label{theorem_ltr-max_pairs_of_pat_312}
Consider the $(312,\sigma)$-machine, where $\sigma=\sigma_1\cdots\sigma_{k-1}\sigma_k\in\mathfrak{S}_k$, with $k\ge 3$ and 
$\sigma_{k-1}<\sigma_k$. Given a permutation $\pi$ of length $n$, let $\pi=M_1B_1\cdots M_tB_t$ be its ltr-max decomposition. Then:
\begin{enumerate}
\item Everytime a ltr-maximum $M_i$ is pushed into the $(312,\sigma)$-stack, the $(312,\sigma)$-stack contains the elements $M_i,M_{i-1},\dots,M_2,M_1$, reading from top to bottom. Moreover, we have
$$
\out^{312,\sigma}(\pi)=\tilde{B}_1\cdots\tilde{B}_tM_t\cdots M_1,
$$
where $\tilde{B}_i$ is a rearrangement of $B_i$.
\item If $\pi$ is $(312,\sigma)$-sortable, then $M_1,M_2,\dots,M_t=n-t+1,n-t+2,\dots,n$.
\end{enumerate}
\end{theorem}
\proof\comm{
1. The proof is identical to that of the first part of Theorem~\ref{theorem_ltr-min_pairs_of_pat_132}.

2.
If $\pi$ is $(312,\sigma)$-sortable, then $\out^{312,\sigma}(\pi)$ avoids $231$ by Proposition 
\ref{KnuthPr}. Suppose, for a contradiction, that there is an element $j\in\lbrace n-t+1,\dots,n\rbrace$ which is not a ltr-maximum. Note that $j\neq\pi_1=M_1$ and $j\neq n=M_t$. Then, $\out^{312,\sigma}(\pi)$ contains an occurrence $jnM_1$ of $231$, which is a contradiction.
}
\endproof

Instantiating $\sigma$ by $123$ in the previous theorem we have the next result.

\begin{theorem}\label{123_312_avoids_213}
Let $\pi$ be a $(123,312)$-sortable permutation and let $\pi=M_1B_1\cdots M_tB_t$ be its ltr-max decomposition. Then:
\begin{enumerate}
\item $B_i$ avoids $213$ for each $i$.
\item $\tilde{B}_i=\out^{12}(B_i)$, for each $i$.
\end{enumerate}
\end{theorem}
\proof\comm{
Let $i\ge 2$. Notice that, as a consequence of Theorem~\ref{theorem_ltr-max_pairs_of_pat_312}, immediately after $M_i$ has been pushed in the $(123,312)$-stack, this stack contains the elements $M_i\cdots M_2M_1$, reading from top to bottom. Moreover, these elements remain at the bottom of the $(123,312)$-stack until the end of the sorting procedure, since they are the last elements of $\out^{123,312}(\pi)$. This fact will be used for the rest of the proof.

1. Suppose, for a contradiction, that $B_i$ contains an occurrence of $213$, for some $i$, and let $bac$ be such an occurrence with $a$ `minimal', in the sense that there is no $a'<a$ where $ba'c$ is an occurrence of $213$ in $B_i$.
Therefore, since $abM_i$ is an occurrence of $123$, $b$ is extracted from the $(123,312)$-stack before $a$ enters. 
In addition, when $c$ enters into the $(123,312)$-stack, $a$ is still in this stack. Indeed, no entry in $B_i$ between 
$a$ and $c$ together with $a$ produces a forbidden pattern in the $(123,312)$-stack. It follows that $\out^{123,312}(\pi)$ contains
$bca$ which is an occurrence of $231$, yielding a contradiction with the sortability of $\pi$.

2. Let us consider the action of the $(123,312)$-stack on the block $B_i$. We wish to show that the behavior of the $(123,312)$-stack when processing $B_i$ is equivalent to the behavior of an empty $12$-stack on input $B_i$. In other words, we prove that the restriction of the $(123,312)$-stack is triggered if and only if the next element of the input forms an occurrence of $12$ together with some other element in the $(123,312)$-stack. Immediately after $M_i$ has been pushed (i.e. before the first element of $B_i$ is processed), the $(123,312)$-stack contains the elements $M_i\cdots M_2M_1$, reading from top to bottom. Observe that $B_i$ avoids $213$ by what proved above, therefore the $(123,312)$-stack cannot be triggered by an occurrence of $312$ when processing $B_i$. Suppose that the next element of the input $x$ forms an occurrence $xy$ of $12$ with some $y\in B_i$. Then $xyM_i$ is an occurrence of $123$ in the $(123,312)$-stack, and so this stack behaves as a $12$-stack.
Conversely, suppose that the $(123,312)$-stack is triggered by an occurrence of $xyz$ of $123$, where $x$ is the next element of the input. Since $M_i>M_{i-1}>\cdots>M_1$, necessarily $y\in B_i$. Thus $xy$ is an occurrence of $12$ that triggers the $12$-stack, as wanted.
}
\endproof

As a consequence of what proved so far in this section, for any $(123,312)$-sortable permutation $\pi=M_1B_1\cdots M_tB_t$ of length $n$, we have $B_i\in\Av(213)$ and $M_1,\dots,M_t=n-t+1,\dots,n$. Moreover, by Proposition~\ref{arrangements}, each $\tilde{B}_i$ in 
$\out^{123,312}(\pi)=\tilde{B}_1\cdots\tilde{B}_tM_t\cdots M_1$ is decreasing. Therefore, for any three elements $x,y,z$, with $x\in B_i$, $y\in B_j$ and $z\in B_k$, with $i<j\le k$, $xyz$ is not an occurrence of $231$. Otherwise $xyz$ would still be an occurrence of $231$ in $\out
^{123,312}(\pi)$, contradicting the fact that $\pi$ is $(123,312)$-sortable. From now on, we say that $xyz$ is an occurrence of $2-3-1$ if $z<x<y$, with $x\in B_i$, $y\in B_j$ and $z\in B_k$, with $i<j<k$. Similarly, when $j=k$, we say that $xyz$ is an occurrence of $2-31$.

\begin{theorem}\label{theorem_4_conditions_123_312}
Let $\pi=M_1B_1\cdots M_tB_t$ be the ltr-max decomposition of a permutation $\pi$ of length $n$. Write $\out^{123,312}(\pi)=\tilde{B}_1\cdots\tilde{B}_tM_t\cdots M_1$ as in Theorem~\ref{theorem_ltr-max_pairs_of_pat_312}. Then $\pi$ is $(123,312)$-sortable if and only if the following conditions are satisfied:
\begin{enumerate}
\item $M_j=n-t+j$, for each $j=1,\dots,t$.
\item $B_i$ avoids $213$ for each $i$ (and thus $\tilde{B}_i$ is decreasing for each $i$).
\item $\out^{123,312}(\pi)$ avoids $2-3-1$.
\item $\out^{123,312}(\pi)$ avoids $2-31$.
\end{enumerate}
\end{theorem}
\proof\comm{
If $\pi$ is $(123,312)$-sortable, then $\pi$ satisfies all the above conditions as a consequence of what proved before in this section. Conversely, it is easy to check that if $\pi$ satisfies the above conditions, then $\out^{123,312}(\pi)$ avoids $231$. Thus $\pi$ is $(123,312)$-sortable.
}
\endproof

Reformulating the third condition of Theorem~\ref{theorem_4_conditions_123_312} we obtain the following lemma, whose easy proof is omitted.

\begin{lemma}\label{lemma_blocks_between_two_el_123_312}
Let $\pi=M_1B_1\cdots M_tB_t$ be the ltr-max decomposition of the $(123,312)$-sortable permutation $\pi$. Write $\out^{123,312}(\pi)=\tilde{B}_1\cdots\tilde{B}_tM_t\cdots M_1$ as in Theorem~\ref{theorem_ltr-max_pairs_of_pat_312}. Then $\out^{123,312}(\pi)$ avoids $2-31$ if and only if for each $x\in B_i$, $y\in B_j$, with $i<j$, we have:
\begin{itemize}
\item if $y>x$, then $B_j>x$.
\item If $y<x$, then $B_j<x$.
\end{itemize}
\end{lemma}

In other words,  Lemma~\ref{lemma_blocks_between_two_el_123_312}
says that each block $B_j$ of a $(123,312)$-sortable permutation $\pi$ is bounded between two previous elements of $\pi$. 
The following result is obtained by restating this lemma and Theorem~\ref{theorem_4_conditions_123_312} in terms of pattern avoidance.

\begin{theorem}
A permutation $\pi$ is $(123, 312)$-sortable if and only if $\pi$ avoids the three generalized patterns 
$[132$, $[42531$ and $[421\overline{5}3$.
\label{bin_patt}
\end{theorem}

Next we prove that $(123,312)$-sortable permutations are enumerated by the binomial transform of Catalan numbers. We shall exploit the above characterization in terms of patterns in order to provide a bijection with a certain set of partial permutations, whose enumeration is straightforward.

Recall from Section~\ref{section_notations} that a partial permutation of $n$ is an injection $\pi:\{1,2,\dots,k\}\to\{1, 2,\dots, n\}$, for some $0\leq k \leq n$. The partial permutation of length zero will be denoted by $\lambda$. Denote by $\mathcal A_n$ the set of all partial permutations of $n$. For instance, we have $\mathcal A_3 = \{\lambda, 1, 2, 3, 12, 21, 13, 31, 23, 32, 123, 132,$ $213, 231, 312, 321\}$. 
There is a natural bijection between the set of permutations in $\mathfrak S_n$ avoiding the pattern $[132$ and $\mathcal A_{n-1}$. Indeed, from a length $n$ permutation $\pi$ avoiding $[132$, we associate the unique partial permutation $\alpha(\pi) \in \mathcal A_{n-1}$
defined as follows:
$$
\alpha(\pi)_{\pi_i}=i-1,\ \mathrm{for}\ \pi_i<\pi_1.
$$

In other words, $\alpha(\pi)$ is obtained by recording the indices (minus one) of the elements $\pi_i<\pi_1$, from the smallest to the largest one. For instance, if $\pi = 52461783$, then $\alpha(\pi) = 4172$. Notice that $\alpha(\pi)=\lambda$ if and only if $\pi_1=1$.
Let us now define two pattern containments on $\mathcal A_n$. Let $a = a_1a_2 \cdots a_m$ be a partial permutation of $n$, with $m \leq n$, and let $i<j<k$. Then $a_ia_ja_k$ is an occurrence of the pattern $31|2$ if it is an occurrence of $312$ such that at least one value of the interval $[a_j,a_k]$ does not appear in $a$. Moreover, we say that $a_ia_ja_k$ is an occurrence of the pattern $\overline{2}1\overline{3}$ if it is an occurrence of $213$ such that $a_i=a_k-1$.\footnote{This is analogous to the notion of bivincular pattern on classical permutations}
By interpreting Theorem~\ref{bin_patt} in terms of partial permutations, we obtain easily:

\begin{theorem}\label{theorem_map_phi}
A permutation $\pi$ is $(123, 312)$-sortable if 
and only if $\alpha(\pi)$ avoids $31|2$ and $\overline{2}1\overline{3}$.
\end{theorem}

Let $\mathcal A_n(31|2,\overline{2}1\overline{3})$ be the set of partial permutations of $n$ avoiding the two patterns 
$31|2$ and $\overline{2}1\overline{3}$, and 
$\mathcal A_n(213)$ be the set of partial permutations of $n$ avoiding the classical pattern $213$.

\begin{figure}[h]
	\label{figPermutation1}
	\begin{center}
			$\pi=~$\scalebox{0.5}	{\begin{tikzpicture}
				\draw [\styleGrille] (\b,\b) -- (\b,\z);
				\draw [\styleGrille] (\d,\b) -- (\d,\z);
                \draw [\styleGrille] (\f,\b) -- (\f,\z);
				\draw [\styleGrille] (\b,\b) -- (\f,\b);

				\draw [\styleGrille] (\b,\d) -- (\f,\d); 
				\draw [\styleGrille] (\b,\g) -- (\f,\g); 
				\draw [\styleGrille] (\b,\z) -- (\f,\z);
                \draw [fill=lightgray](\b,\b)--(\b,\d)--(\d,\d)--(\d,\b); \draw [fill=lightgray](\d,\B)--(\f,\B)--(\f,\b)--(\b,\b);
                \draw [fill=lightgray](\d,\d)--(\f,\d)--(\f,\g)--(\d,\g)--(\d,\d);
                \draw [fill=lightgray](\b,\g)--(\b,\z)--(\d,\z)--(\d,\g);
                \draw [\styleGrille] (\b,\B) -- (\f,\B);
                \draw[fill=lightgray](\b,\g)--(\b,6.25cm)--(\f,6.25cm)--(\f,\g);
               \point{\d}{\B};\draw (\c,\E) node {\Large $A$};
               \draw (\e,2.25cm) node {\Large $B$};
               \draw (\e,7.25cm) node {\Large $B$};
               \draw (\a,6.1cm) node {\Large $x$};

			\end{tikzpicture}}
$\longrightarrow\pi'=~$\scalebox{0.5}	{\begin{tikzpicture}
				\draw [\styleGrille] (\b,\b) -- (\b,\z);
				\draw [\styleGrille] (\d,\b) -- (\d,\z);
                \draw [\styleGrille] (\f,\b) -- (\f,\z);
				\draw [\styleGrille] (\b,\b) -- (\f,\b);
				\draw [\styleGrille] (\b,\d) -- (\f,\d); 
\draw [\styleGrille] (\b,3.25cm) -- (\f,3.25cm); 
				\draw (\g,3cm) node {\Large $r+1$};
				\draw [\styleGrille] (\b,\z) -- (\f,\z);
                \draw [fill=lightgray](\d,5cm)--(\d,\z)--(\f,\z)--(\f,5cm);
                \draw [fill=lightgray](\b,\b)--(\b,5cm)--(\d,5cm)--(\d,\b);
                \draw [fill=lightgray](\b,\d)--(\b,3.25cm)--(\f,3.25cm)--(\f,\d);\draw [fill=lightgray](\d,\B)--(\f,\B)--(\f,\b)--(\b,\b);

				\draw [\styleGrille] (\b,\d) -- (\f,\d); 
                \draw [\styleGrille] (\b,5cm) -- (\f,5cm);
                \draw [\styleGrille] (\b,3.25cm) -- (\f,3.25cm); 
               \point{\d}{\B};

               \draw (\e,2.25cm) node {\Large $B'$};
               \draw (\e,4.25cm) node {\Large $B'$};
               \draw [\styleGrille] (\b,\b) -- (\f,\b);
				\draw [\styleGrille] (\b,\d) -- (\f,\d); 
\draw [\styleGrille] (\b,3.25cm) -- (\f,3.25cm); 
				
              \draw [\styleGrille] (\b,\B) -- (\f,\B);
				\draw [\styleGrille] (\b,\z) -- (\f,\z);
                \draw (\c,\G) node {\Large $A'$};
			\end{tikzpicture}}
$\mbox{ and }\phi(\pi)=~$\scalebox{0.5}	{\begin{tikzpicture}
				\draw [\styleGrille] (\b,\b) -- (\b,\z);
				\draw [\styleGrille] (\d,\b) -- (\d,\z);
                \draw [\styleGrille] (\f,\b) -- (\f,\z);
				\draw [\styleGrille] (\b,\b) -- (\f,\b);
				\draw (\G,2.75cm) node {\Large $y<=r+1$};
				\draw [\styleGrille] (\b,\z) -- (\f,\z);
                \draw [fill=lightgray](\d,5cm)--(\d,\z)--(\f,\z)--(\f,5cm);
                \draw [fill=lightgray](\b,\b)--(\b,5cm)--(\d,5cm)--(\d,\b);
                \draw [fill=lightgray](\b,\C)--(\b,2.75cm)--(\f,2.75cm)--(\f,\C);
                \draw [fill=lightgray](\d,\B)--(\f,\B)--(\f,\b)--(\b,\b);

				\draw [\styleGrille] (\b,\C) -- (\f,\C); 
                \draw [\styleGrille] (\b,5cm) -- (\f,5cm);
                \draw [\styleGrille] (\b,2.75cm) -- (\f,2.75cm); 
               \point{\d}{\B};

               \draw (\e,2cm) node {\Large $\phi(B')$};
               \draw (\e,4cm) node {\Large $\phi(B')$};
               \draw [\styleGrille] (\b,\b) -- (\f,\b);
				
              \draw [\styleGrille] (\b,\B) -- (\f,\B);
				\draw [\styleGrille] (\b,\z) -- (\f,\z);
                \draw (\c,\G) node {\Large $\phi(A)$};
			\end{tikzpicture}}

		\end{center}
	\caption{Illustration of $\phi$ in the case ($iii$) of the proof of Theorem~\ref{theorem_map_phi}.}\label{figure_map_phi}
\end{figure}
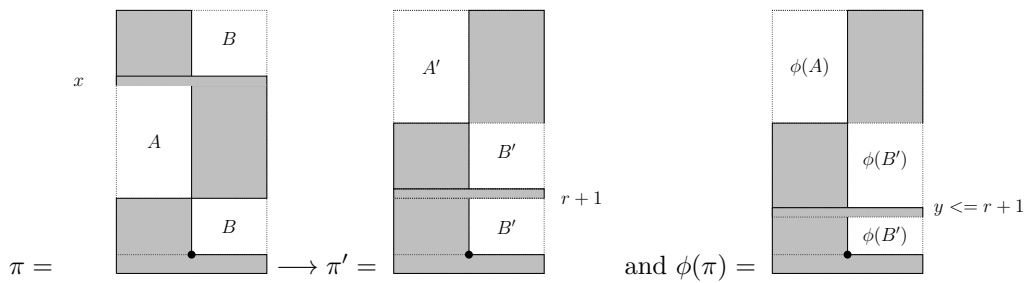

\begin{theorem}
For any $n\geq 1$, there is a bijection $\phi$ between $\mathcal A_n(31|2,\overline{2}1\overline{3})$ and $\mathcal A_n(213)$.
\end{theorem}
\proof Let us define recursively the map $\phi$ from $\mathcal A_n(31|2,\overline{2}1\overline{3})$ to $\mathcal{A}_n(213)$. If $\pi=\lambda$, then we set $\phi(\pi)=\lambda$. Otherwise, $\pi$ has a unique decomposition of the form $\pi=A \min(\pi) B$ where $A$ and $B$ are disjoint partial permutations of $n$. We distinguish three cases:
\begin{itemize}
\item[($i$)] If at least one of $A$ or $B$ is empty, then we set $\phi(\pi)=\phi(A)\min(\pi)\phi(B)$;
\item [($ii$)] If both $A$ and $B$ are not empty and  $\min(A)>\max(B)$, then we set $\phi(\pi)=\phi(A)\min(\pi)\phi(B)$. It is worth noting that the hypothesis that $\pi$ avoids $31|2$ implies that any value $x\in [\min(\pi),\max(B)]$ occurs in $B$.

\item [($iii$)] Suppose that both $A$ and $B$ are not empty and $\min(A)<\max(B)$. Since $\pi$ avoids $\overline{2}1\overline{3}$,  there exists $x\in [\min(A),\max(B)]$ such that $x$ does not occur in $\pi$.  We choose the smallest $x$ with this property, so that any value of the interval $[\min(A),x]$ occurs in $A$. Moreover, since $\pi$ avoids  $31|2$, it must be $\max(A)=x-1$. An illustration of this case is depicted in Figure~\ref{figure_map_phi}. 
Let $r$ be the maximum value of $B$ that is lower than $\min(A)$ and consider the string $B'$ obtained from $B$ by decreasing by $x-r-1$ all entries greater than $x$. Similarly, let $A'$ be obtained from $A$ by increasing by $\max(B)-x+1$ all its entries. Obviously, $A'$ and $B'$ belong to $\mathcal{A}_n(31|2,\overline{2}1\overline{3})$, whereas $\pi'=A'\min(\pi)B'$ contains $31|2$. 
Then we set $\phi(\pi)=\phi(A')\min(\pi)\phi(B')$ (see again Figure~\ref{figure_map_phi} for an illustration of this mapping). It is worth noting that the value $r+1$ does not occur in both $A'$ and $B'$, which implies that there exists $y\in [\min(\pi),r+1]$ such that $y$ does not occur in $\phi(B')$.
\end{itemize}

Next we prove that $\phi$ is an injective map. We proceed by induction on the length of partial permutations. Let $\pi$ be a partial permutation. Due to the remarks at the end of ($ii$) and ($iii$), the image of $\pi$ under $\phi$  satisfying ($ii$) is a partial permutation $\pi'$ such that   any value $x\in [\min(\pi),\max(\phi(B))]$  occurs in $\phi(B)$, which is not true for a permutation $\pi$ satisfying ($iii$). Then, for two non-empty partial permutations $\pi$ and $\sigma$ in $\mathcal A_n(31|2,\overline{2}1\overline{3})$, $\phi(\pi)=\phi(\sigma)$  implies that $\pi$ and $\sigma$ have the same length and they belong to the same case ($i$), ($ii$) or ($iii$). The recurrence hypothesis induces $\pi=\sigma$ which completes the induction.

Finally, observe that any partial permutation $\pi$ avoiding $213$ is of the form $A\min(\pi)B$ where $\min(A)>\max(B)$ and both $A$ and $B$ avoid $213$. According to the geometrical shape of $\pi$ (as in the proof of injectivity), $\pi$ fits exactly in one of the cases ($i$), ($ii$) and ($iii$) in the definition of $\phi$. Therefore the surjectivity of $\phi$ can be showed by using its recursive definition and induction on $A$ and $B$. We leave the details to the reader.
\endproof

Now, it is easy to enumerate the set $\mathcal{A}_n(213)$. Indeed any partial permutation $\pi\in \mathcal{A}_n(213)$ can be obtained by choosing $k$ integers from $\lbrace 1,2,\dots,n\rbrace$ and then arranging them according to the partial order of a permutation in $\Av_k(213)$ (there are $c_k$ such permutations). Therefore, we have:
$$|\mathcal{A}_n(213)| = \sum_{k=0}^{n} \binom{n}{k}c_k,$$
the binomial transform of Catalan numbers (sequence {\tt A007317} in~\cite{Sloane}). The enumeration of $\Sort(123,312)$ follows immediately.

\begin{corollary}
For each $n\ge 1$, we have:
$$|\Sort_{n+1}(123, 312)| = \sum_{k=0}^{n} \binom{n}{k}c_k.$$
\end{corollary}

\end{document}